\documentclass[aps,superscriptaddress,twocolumn,prl]{revtex4}
\usepackage{epsfig,graphics,amssymb,amsmath,subeqnarray,amsthm,latexsym,times}
\def\d{{\rm d}}
\def\r{{\bf r}}\def\p{\partial}\def\p{{\bf {p}}}

\def\Re{{\rm Re}}\def\Ca{{\rm Ca}}\def\Me{{\rm Me}}\def\El{{\rm El}}
\def\M{{\bf M}}\def\e{{\bf e}}\def\q{{\bf q}}\def\g{ \gamma}
\def\N{{\bf N}}
\begin{document}
\title{Soft swimming: Exploiting deformable interfaces for low-Reynolds number  locomotion}
\author{Renaud Trouilloud}
\affiliation{
Hatsopoulos Microfluids Laboratory, Department of Mechanical Engineering,
Massachusetts Institute of Technology,
77 Mass. Ave.,
Cambridge, MA 02139, USA.}
\author{Tony S. Yu}
\author{A. E. Hosoi}
\affiliation{
Hatsopoulos Microfluids Laboratory, Department of Mechanical Engineering,
Massachusetts Institute of Technology,
77 Mass. Ave., 
Cambridge, MA 02139, USA.}
\author{Eric Lauga$^*$
}
\affiliation{
Department of Mechanical and Aerospace Engineering, University of California San Diego,\\ 9500 Gillman Dr., La Jolla CA 92093-0411, USA.\\
$^*${\rm Corresponding author (elauga@ucsd.edu)}
}
\date{\today}
\begin{abstract}

Reciprocal movement cannot be used for locomotion at low-Reynolds number in an infinite fluid or near a rigid surface.
Here we show that this limitation is relaxed for a body performing  reciprocal motions near a deformable interface. 
Using physical arguments and scaling relationships, we show that the nonlinearities arising from reciprocal flow-induced interfacial deformation rectify the periodic motion of the swimmer, leading to locomotion. Such a strategy can be used to move toward, away from, and parallel to any deformable interface as long as the length scales involved are smaller than intrinsic scales, which we identify. A macro-scale experiment of flapping motion near a free surface illustrates this new result.

\end{abstract} 
\maketitle

Swimming microorganisms inhabit a world quite different from the one we experience.
Their motion through the surrounding fluid occurs at very low Reynolds numbers ($\Re$), a fact with two important consequences: (a) the only physical force available to produce thrust is drag; (b) since the fluid equations  (Stokes equations) are linear and time-reversible, swimming motions symmetric with respect to time reversal (reciprocal motion) cannot be used for locomotion (the scallop theorem \cite{purcell77}). Biological swimmers, such as bacteria and spermatozoa, overcome these two limitations by exploiting the anisotropic drag of slender filaments such as flagella and cilia, and  actuating these filaments in  a wave-like fashion \cite{lighthill76,braybook}.

Motivated by the recent development of artificial swimmers \cite{dreyfus05_nature}, we pose here the following general question: are there any new low-$\Re$ swimming methods remaining to be discovered? In particular, can some simple reciprocal movements produce locomotion, thereby apparently violating the constraints of the scallop theorem?

In this paper, we propose a new method for locomotion without inertia.  Using theory and  experiments, we  show that  a body performing a reciprocal movement is able to move near a soft interface: physically, the deformation of the interface provides the geometric nonlinearities necessary to escape the constraints of the scallop theorem.
This strategy, which is relevant for  sufficiently small systems, also implies that the time-reversible component of all swimmers (including biological organisms) can generate  non-trivial flows and propulsive forces near soft interfaces.

\begin{figure}[b]
\centering
\includegraphics[width=.44\textwidth]{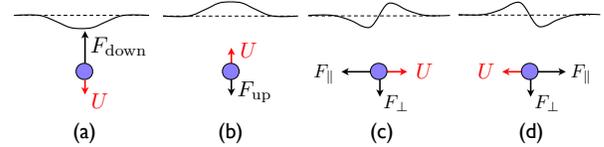}
\caption{Breakdown of reversibility near soft interfaces: a sphere moving away from a soft surface 
 experiences more drag than one moving toward  the interface, $F_{\rm down} > F_{\rm up}$  (a, b); 
a  sphere translating parallel to a soft surface experiences a lift force, $F_\perp$, perpendicular to the direction of motion (c, d).}
\label{deform_model}
\end{figure}

The physical picture for soft swimming arises from the asymmetries between different modes of motion near deformable interfaces, and can be  illustrated by considering the  motion of a rigid sphere slowly translating below a free surface (Fig.~\ref{deform_model}).  For motion perpendicular to the interface, the drag on the sphere is always increased by the presence of the interface; however, the magnitude and direction of the surface-induced component of the drag can be modulated by deforming the interface \cite{berdan82,leal82}. 
When the sphere moves away from the interface (Fig.~\ref{deform_model}a), it displaces fluid with it, and the interface deforms toward the sphere, resulting in an increase of the surface-induced drag, $F_{{\rm down}}>F_{\rm flat}$. 
In contrast, when the sphere moves toward the interface (Fig.~\ref{deform_model}b), the free surface is  displaced  away from the sphere, and the component of the drag due to the interface is diminished, $F_{{\rm up}}<F_{\rm flat}$. The asymmetry, $\Delta F =F_{{\rm down}}-F_{{\rm up}} > 0$, results in an averaged net force on the fluid, $\Delta F$, away from the interface if the sphere is oscillated in a reciprocal fashion. (Alternatively, if  the sphere is acted upon by an external force with zero average, it will move toward the interface). 
For motion  parallel to the interface (Figs.~\ref{deform_model}c and d), surface deformation leads to a lift  force on the sphere, perpendicular to its direction of motion and directed away from the interface \cite{beaucourt04,skotheim04}, reminiscent of the lift force responsible for the migration of deformable bodies such as droplets \cite{leal80}, vesicles \cite{cantat99,seifert99} and polymers \cite{agarwal94}  away from boundaries. 
These effects are  due  to the nonlinear (geometric) coupling between the moving sphere and the deformable interface, and they vanish when the interfaces do not deform. A similar coupling can be exploited for locomotion.

\begin{figure}[t]
\centering
\includegraphics[width=.44\textwidth]{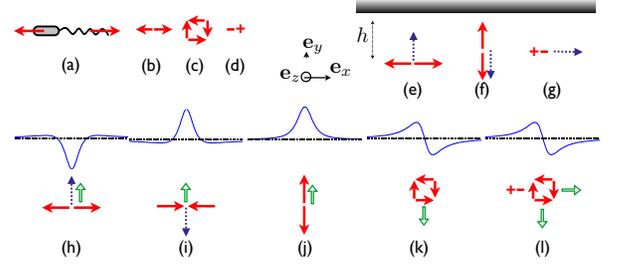}
\caption{Physics of soft swimming.
(a) The lowest-order flow around a low-$\Re$ swimmer  decays as a force-dipole;
(b): parallel force-dipole ($p_{xx}>0$);
(c): perpendicular force-dipole ($p_{xy}<0$);
(d):  potential  (source) dipole ($q_x>0$);
(e): $p_{xx}>0$ is attracted by a flat surface;
(f): $p_{yy}>0$ is repelled by a flat surface;
(g): $q_x<0$ moves parallel to a flat surface;
(h): $p_{xx}>0$ deforms a soft surface in a manner that enhances the attraction;
(i):  $p_{xx}<0$ deforms the same surface in a manner that reduces the repulsion;
(j): a reciprocal $p_{yy}$ is attracted by a soft surface;
(k): a reciprocal $p_{xy}$ is repelled by a soft surface;
(l):  a reciprocal  combination of $p_{xy}$ and $q_x$ moves away from, and parallel to a deformable surface.}
\label{soft_swimming}
\end{figure}

Let us consider a  body - a swimmer -  that deforms in a prescribed and periodic fashion.  On length scales larger than the typical size of the swimmer, the disturbance flow field can be modeled as  a superposition of point flow disturbances. Since a low-$\Re$ swimmer cannot exert any net force on the surrounding fluid, the instantaneous velocity field decays at most like a force-dipole (Fig.~\ref{soft_swimming}a); the swimmer is also torque-free, and therefore only symmetric force dipoles  are allowed (stresslets \cite{batchelor70,pedley92}). In three-dimensions, the dipole strength, ${\bf p}$, is a 3-by-3 tensor with 6 independent components. The associated flow field decays as $\sim 1/r^2$ and is given by
${\bf u}^p=\frac{p_{ij}}{8\pi\eta r^3}\left(-\delta_{ij}+3\frac{x_ix_j}{r^2} \right){\bf r}$,
where $\eta$ is the fluid viscosity, and $\r = x_i\e_i$ is the distance to the dipole.
There exist two types of dipoles: parallel dipoles, in which the direction of the forces is aligned with that of the force gradient  (Fig.~\ref{soft_swimming}b), and perpendicular dipoles, in which  these directions are perpendicular to one another (Fig.~\ref{soft_swimming}c). Furthermore, in general, the flow field can contain other types of singularities (higher moments  of force distribution), with faster spatial decay, predominantly the potential (source) dipole (Fig.~\ref{soft_swimming}d), of vector strength $\q$, associated with a velocity field decaying as $\sim 1/r^3$  and given by 
$u^q_i=\frac{q_{j}}{8\pi\eta r^3}\left(-\delta_{ij}+3 \frac{x_ix_j}{r^2} \right)$.

In an infinite fluid, and by linearity of Stokes equations, the velocity of a swimmer whose far field is described by a superposition of stresslets and potential dipoles is given by  ${\bf u}={\M}\cdot {\bf p} + {\N}\cdot {\bf q}$, where the mobility matrices $\M$ and $\N$ are functions of the instantaneous shape of the swimmer; here, we will take $\M$ and $\N$ as constants, a suitable approximation for bodies performing small-amplitude swimming motions. We also consider swimmers with sufficiently symmetric motion so that their orientation remains constant and ignore torque balance. 

We now  examine the case of a swimmer located near an undeformable interface. 
The presence of the interface will result in additional velocity for the swimmer, ${\bf u}_w (h)$, where $h$ is the distance to the interface, due to the image singularities required  to enforce the correct boundary conditions at the interface \cite{blake74}. The equation for the swimming velocity becomes
${\bf u} = {\bf M}\cdot \p+ {\N}\cdot \q + {\bf u}_w(h)$, where the wall-induced velocity is given by
${\bf u}_w(h)  =  \frac{1}{32\pi\eta h^2}[(p_{xx}+p_{zz}-2p_{yy}){\bf e}_y-\frac{1}{2h}(q_x\e_x+q_z\e_z + 2 q_y \e_y)]$ in the case of a flat free surface
(illustrated by the dashed  arrows in Figs.~\ref{soft_swimming}e-g);
results for no-slip surfaces are similar. Note that the magnitude of ${\bf u}_w (h)$  scales linearly with the dipole strengths, $\p$ and $\q$. In general, surfaces of fixed shape will lead to an equation for the swimming kinematics given by
$\frac{\d \r }{ \d t} = {\bf A}({\bf r}) \cdot \p(t) + {\bf B}({\bf r}) \cdot \q(t).$
For reciprocal deformations, we have $\p(t)=\p_0 f(t)$, $\q(t)=\q_0 f(t)$, where $f$ is an arbitrary periodic function of time: the swimming kinematics are described by  an autonomous dynamical system, and no locomotion can occur 
 on average, $\langle \bf u \rangle = 0$. This is the scallop theorem.

We now consider  interfaces that can deform due to the flow set up by the swimmer. Since wall effects depend on the shape of, and the distance to the interface, and since both can be modified by the swimmer's flow field, the relationship between the local reciprocal actuation and  the swimming kinematics becomes nonlinear. 

Let us  consider a free surface, with surface tension $\g$, and show that motion toward, away from, and parallel to the surface can be devised. For motion toward the surface, we only need to consider parallel force-dipoles aligned with the surface, $\p= p(t) \e_x\e_x$ (Figs.~\ref{soft_swimming}h-i). We assume  small Capillary number to capture the first-order effect of surface deformation ($p/\gamma h^2\ll 1$). The normal deformation of the interface (magnitude $\delta$) is set by the balance between capillary  and viscous forces at the interface, $\gamma \delta  /h^2 \sim - p/h^3$, hence $\delta  \sim - p/\gamma h$; note that the  surface deformation can be calculated analytically for small Capillary numbers, as plotted in Figs.~\ref{soft_swimming}h-l.
The wall-induced velocity on the dipole, due to the image system, now has a magnitude 
$u_w =\e_y \cdot {\bf u}_w \sim p/\eta (h+\delta)^2$, and can be linearized around $h_0$ (the starting position of the swimmer) to obtain 
$u_w  \sim p/\eta (h_0)^2  -p\delta/\eta h_0^3 \sim p/\eta (h_0)^2  +p^2/\gamma \eta h_0^4$. The swimming kinematics in the $y$ direction are then given by ${\d h}/ {\d t} \sim  Mp(t) +{p(t)^2}/{\gamma \eta h_0^4} -{p(t)}/{\eta h_0^2}$. 
For reciprocal swimming $p(t)=p_0\cos \omega t$, we get a net locomotion toward the surface, with average swimming velocity
$\langle u_y \rangle \sim  {p_0^2}/{\gamma \eta h_0^4}$
\footnote{The linearization around $h_0$ is valid when  $Mp_0/\omega h_0 \ll 1$ and $p_0/\eta h_0^3\omega \ll 1$; the quasi-steady approximation for the surface deformation is valid for   $\omega \ll \eta/ \rho h_0^2$.}.  In practice, the dipole arises from the unsteady actuation of a swimmer of size $L$ with velocity $U\sim \omega L$ , and therefore  $p_0\sim\eta L^2 U$, so that $\langle u_y \rangle /U \sim (\eta U/ \gamma) (L/h_0)^4$: for sufficiently large deformations $\eta U/ \gamma \sim 1$ and confined swimmers $L/h_0\sim 1$, we obtain a net speed on the order of the  actuation speed
$\langle u_y \rangle \sim U$ \footnote{Interestingly, for  real spermatozoa, the swimming speed is only a fraction of the actuation speed of its flagellum (swimming at about $100-200$ $\mu$m/s with a  $30-60$ $\mu$m-long  flagella oscillating at $30-50$ Hz); hence,  even for weakly confined systems, the effect from soft swimming could be as important as the ``natural'' swimming speed.}.

The physical interpretation for this motion toward the surface is illustrated in Figs.~\ref{soft_swimming}h-i. When $p>0$, the swimmer deflects the surface toward it, enhancing the wall-induced attraction (Fig.~\ref{soft_swimming}h, dashed arrow) by a small amount (empty arrow); when $p<0$, the surface is deflected away from the swimmer, reducing the wall induced repulsion (Fig.~\ref{soft_swimming}i, dashed arrow), by a small amount (empty arrow). Both deformation-induced effects are directed toward the surface, and therefore do not average to zero for a reciprocal actuation. The results are similar for a parallel force dipole in the direction perpendicular to the surface, $\p= p(t) \e_y\e_y$ (Fig.~\ref{soft_swimming}j).

Locomotion away from the surface can be obtained by considering perpendicular force dipoles, of the form $\p= p(t) (\e_x\e_y+\e_y\e_x)$ (Fig.~\ref{soft_swimming}k). In that case, the surface deformation is asymmetric, and it induces a lift force which does not average to zero with $ p$ \cite{berdan82}. The scaling for the time-averaged swimming speed is similar to the previous section, except that it is now directed away from the free surface, 
$\langle u_y \rangle \sim -  {p_0^2}/{\gamma \eta h_0^4}$.

Moving parallel to the deformable surface is more subtle. Indeed, since the first effect of surface deformation is always directed perpendicular to the undisturbed surface \cite{berdan82}, the only way to move parallel to the surface is to use motion toward or away from the surface to rectify an otherwise time-periodic and spatially dependent parallel motion (such as the one illustrated in Fig.~\ref{soft_swimming}g).  
This can be achieved, for example,  by combining a  perpendicular force dipole, $\p= p(t)  (\e_x\e_y+\e_y\e_x)$, with a potential dipole, $\q  = q(t) \e_x$ (Fig.~\ref{soft_swimming}l). Since the flow field due to the potential dipole decays faster (by one order of magnitude) than that due to the force-dipole, the first effect of surface deformation will be the establishment of the lift force due to $p$. The swimming kinematics are therefore given by ${\d h}/ {\d t} \sim -{p(t)^2}/{\gamma \eta h^4} $ in the direction perpendicular to the surface, and 
${\d x}/ {\d t} \sim  - {q(t)}/{\eta h^3}$ in the parallel direction. Writing $h(t)= h_0 + \tilde h(t)$, with $\tilde h(t)\ll h_0$, we have 
${\d x}/ {\d t} \sim -{q}/{\eta h_0^3} + {q\tilde h}/{\eta h_0^4}$. For a reciprocal motion with $p(t) = p_0\cos (\omega t + \varphi)$ and $q(t) = q_0\cos (\omega t + \varphi)$, time integration leads to
$\tilde h (t) = -p_0^2(t + \sin(2\omega t + 2\varphi)/2\omega)/\gamma\eta h_0^4 $, and therefore the average swimming speed in the $x$ direction is given by  $\langle u_x \rangle \sim -  {p_0^2q_0 \sin \varphi}/\omega {\gamma \eta^2 h_0^8}$. 
Thus swimming occurs both away from the surface and in the $+$ or $-x$ direction, depending on the sign of $\sin \varphi$.  Similarly, combining  a force dipole in the  $\e_x\e_x$
or  $\e_y\e_y$ direction with a  source dipole along $\e_x$ leads to combined motion toward  and parallel to  the surface.

\begin{figure}[t]
\centering
\includegraphics[width=.44\textwidth]{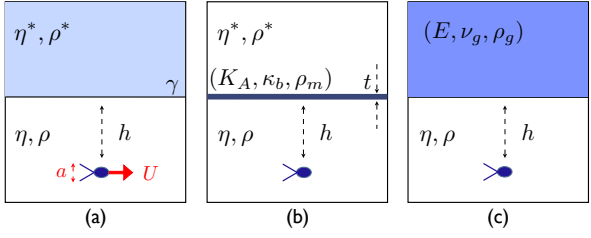}
\caption{Examples where soft swimming can occur:  (a) interface between two liquids (or between a liquid and a gas), (b) elastic membrane separating two liquids, and (c) interface between a liquid and an elastic gel (see text for notations).}
\label{three}
\end{figure}

The examples above demonstrate that a free surface can be exploited to obtain locomotion in all directions using reciprocal forcing. To analyze the effectiveness of these swimming strategies, three types of biologically-relevant soft surfaces can be considered (Fig.~\ref{three}): the interfaces between two liquids (Fig.~\ref{three}a), an elastic membrane separating two liquids (Fig.~\ref{three}b) and the interface between a liquid and an elastic gel (Fig.~\ref{three}c).

Consider a fluid with viscosity $\eta$ and density $\rho$,  in which all length scales $L$ (swimmer size, $a$, distance to the surface, $h$) are similar. For soft swimming to occur, the local reciprocal actuation of the swimmer has to generate flow speeds, $U$, sufficiently small for the Reynolds number, $\Re= \rho U L/ \eta$, to be small, yet large enough to induce significant deformation of the interface, {\it i.e.} a large (a) Capillary number, $\Ca= \eta U / \g$,  for the  interface between two fluids with similar viscosities ($\eta$, $\eta^*$) and densities ($\rho$, $\rho^*$),
(b) dimensionless  membrane-viscous number, $\Me = \eta U / K_A$, for two fluids separated by an elastic 
membrane (thickness $t$, area modulus $K_A$, density $\rho_m$, bending stiffness $\kappa_B$), 
(c) elasto-viscous number, $\El = \eta U / L E$, for an elastic gel (Young's modulus $E$, Poisson's ratio, $\nu_g$, density $\rho_g$). By combining 
$\Re <1$ with 
(a) $\Ca > 1$ , 
(b)  $\Me>1$,
(c) $\El>1$, we find that soft swimming is expected to be effective for length scales below 
(a) $L<\ell_f$,  where $\ell_f = \eta^2/\rho\gamma$ is an intrinsic visco-capillary length scale (the Ohnesorge length), 
(b) $L<\ell_m$, where $ \ell_m =  {\eta^2}/{\rho  K_A}$ is the intrinsic membrane-viscous length scale,
(c) $ L <  \ell_g $, where $  \ell_g =  \sqrt{{ \eta^2}/{\rho E}}$ is the intrinsic elasto-viscous length scale
 \footnote{In order for gravity to be irrelevant, the length scales should be small enough for the appropriate Bond numbers to be small; this is usually satisfied in biologically-relevant situations where  density differences are negligible.}.

In order to probe the biological relevance of these ideas, we consider an environment where motility is crucial, namely the female reproductive tract in humans \cite{suarez06}. In this case, the fluids involved (e.g. cervical mucus) have viscosities in the range  $\eta \sim 0.1-10$ Pa.s \cite{wolf77_1}, leading to values of $\ell_f$ between fractions of millimeters to meters. A similar range for $\ell_m $ is obtained in the case where such fluids are separated by a  lipid bilayer \cite{boal_book}, and for $\ell_g$ when the fluids are near an elastic gel such as a cross-linked actin network \cite{boal_book}. In all cases,  the intrinsic length scales are large, and therefore biological flows  arising from time-reversible motion could be substantial.

\begin{figure}[t]
\centering
\includegraphics[width=.43\textwidth]{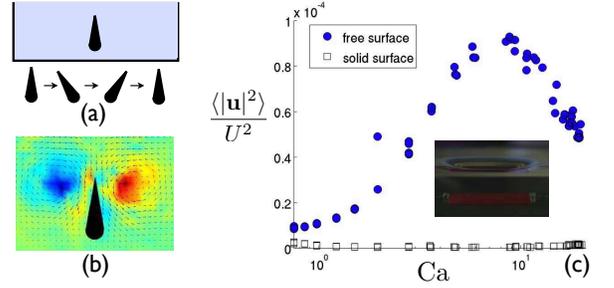}
\caption{
Experiment to generate a net flow by reciprocal flapping motion below a free surface;  
(a) principle of the experiment; 
(b) velocity and vorticity fields over one  flapping cycle (red is positive vorticity);
(c) dimensionless kinetic energy over one  cycle as a function of the flapping Capillary number, $\Ca$ (insert, photograph of the free surface deformation).}
\label{flapper_data}
\end{figure}

To further illustrate the breaking of the scallop theorem near soft surfaces, we exploit the increase of the intrinsic scale, $\ell_f$, with the viscosity,  and perform a macro-scale experiment near a free surface. Instead of a swimmer, we consider a flapper fixed in place, for which the problem of reciprocal swimming is replaced by the dual problem of reciprocal pumping, {\it i.e.} creating a non-zero flow from a time-periodic forcing. 

A thin plate, the flapper (size 2~cm $\times$ 7~cm), was mounted to the shaft of a servo motor and rotated  over a range of 90$^\circ$ (45$^\circ$ on each side) at flapping periods of 3--200~s (Fig.~\ref{flapper_data}a). The flapper was immersed in a high-viscosity silicone oil ($\eta =60$~Pa$\cdot$s, $\ell_f\approx 50$~m) with the tip of the flapper positioned 2~cm below the free surface; alternatively, the free surface  could be covered by a solid acrylic sheet to prevent surface deformation. A laser sheet illuminated  polyamide beads  (diameter, 93~$\mu m$) dispersed in the oil, and images of the beads were captured after every flapping cycle using a digital SLR camera with 3072 $\times$ 2048 pixels per image. The displacement field per flapping cycle was calculated from sequential images using the MatPIV toolbox for Matlab.

The experimental results are presented in Fig.~\ref{flapper_data}b-c. When the surface is free to deform, the reciprocal flapping motion induces  a net flow. An example of the vorticity contours  (red is positive vorticity) and flow field over one flapping cycle
is displayed in Fig.~\ref{flapper_data}b: the net flow is induced along the flapper, away from the free surface;  when the free surface is replaced by a solid lid, this flow disappears.  Measurements of the kinetic energy in the flow (integrated over the whole system) are shown in Fig.~\ref{flapper_data}c (nondimensionalized by flapper length and flapping frequency). When the  surface is solid (hollow symbols), no flow occurs on average, whereas when the surface is free to deform, the average flow possesses finite kinetic energy (filled symbols),  demonstrating that reciprocal actuation near a soft surface can  be used to escape the constraints of the scallop theorem.

In summary, we have shown that soft interfaces allow propulsion driven by reciprocal motion. This new strategy for low-Reynolds number swimming means not only that primitive swimming methods become effective near a soft interface, but more generally that the  usually-neglected time-reversible component of all swimmers 
(including biological organisms and appendages)  can generate  non-trivial flows and propulsive forces in such settings. The  symmetry-breaking arises from the normal stress boundary condition: changing the sign of the forcing by a reciprocal swimmer changes the sign of the interfacial normal stresses, thereby changing the interface curvature, and its shape.  The strategy is relevant for sufficiently small systems. 
Moreover, for swimmers of size $L$ at a distance $h$ from a surface, the swimming efficiency is expected to scale as ${\cal E}\sim (L/h)^n$ ($n>1$) and will therefore be most efficient for confined geometries where $L\sim h$, for example for the motion of swimmers in thin films and membranes, or near surfaces. Similar results are expected for the dual problem of generating feeding currents from reciprocal forcing. Possible extensions of this work include investigating surface-induced reorientation,  rotational swimming, and a closer examination of cases in which  surface deformations are large. Together with the current work, this would provide a comprehensive framework for understanding the consequence of soft interfaces on locomotion and transport on small scales.


We thank Denis Bartolo for useful discussions. 
This work was funded by the NSF (grant CTS-0624830).

\bibliographystyle{prsty}
\bibliography{biblio_deformable_interface}
\end{document}